\documentclass[12pt]{article}
\usepackage[a4paper, total={6.3in, 9.7in}]{geometry}
\usepackage[utf8]{inputenc}
\usepackage[english]{babel}
\usepackage{blindtext}
\usepackage{amsmath}
\usepackage{physics}

\usepackage{xcolor}
\usepackage{soul}
\usepackage[T1]{fontenc} 
\usepackage{graphicx}
\usepackage{authblk}
\usepackage[backend=biber, sorting=none, style=nature]{biblatex}
\usepackage[compact]{titlesec}
\usepackage{float}

\usepackage[font=footnotesize,labelfont=bf]{caption}

\usepackage[hypertexnames=false,colorlinks=true,allcolors=blue]{hyperref}
\DeclareUnicodeCharacter{0301}{************}

\newcommand{\beginsupplement}{%
	\setcounter{table}{0}
	\renewcommand{\thetable}{S\arabic{table}}%
	\setcounter{figure}{0}
	\renewcommand{\thefigure}{S\arabic{figure}}%
	\setcounter{section}{0}
	\renewcommand{\thesection}{S\arabic{section}}
	\setcounter{equation}{0}
	\renewcommand{\theequation}{S\arabic{equation}}
}

\title{Mobility enhancement in CVD-grown monolayer $\textrm{MoS}_2$ via patterned substrate induced non-uniform straining}

\author[1,2]{Arijit Kayal\footnote{Contributed equally}}

\newcommand\CoAuthorMark{\footnotemark[\arabic{footnote}]} 
\author[1]{Sraboni Dey\protect\CoAuthorMark}
\author[1]{Harikrishnan G.} 
\author[1]{Renjith Nadarajan}
\author[1]{Shashwata Chattopadhyay}
\author[1,3]{J. Mitra}

\affil[1]{%
	\textit{School of Physics, Indian Institute of Science Education and Research, Thiruvananthapuram, Kerala 695551, India}}
\affil[2]{arijit17@iisertvm.ac.in}
\affil[3]{j.mitra@iisertvm.ac.in}

\date{}
\begin{document}
	\maketitle
	\begin{abstract}
		The extraordinary mechanical properties of 2D TMDCs  make them ideal candidates for investigating strain-induced control of various physical properties. Here we explore the role of non-uniform strain in modulating optical, electronic and transport  properties of semiconducting, chemical vapour deposited monolayer $\textrm{MoS}_2$, on periodically nanostructured substrates. A combination of spatially resolved spectroscopic and electronic properties explore and quantify the differential strain distribution and carrier density on a monolayer, as it conformally drapes over the periodic nanostructures. The observed accumulation in electron density at the strained regions is supported by theoretical calculations which form the likely basis for the ensuing $\times60$ increase in field effect mobility  in strained samples. Though spatially non-uniform, the pattern induced strain is shown to be readily controlled by changing the periodicity of the nanostructures thus providing a robust yet useful macroscopic control on strain and mobility in these systems.
	\end{abstract}
\begin{center}
\end{center} 
	\section{Introduction}
\label{Introduction}
Both natural and synthetic two-dimensional (2D) materials have continued to dominate the area of materials research owing to a multitude of fascinating physical properties they exhibit. The relevance of individual materials stem not only from niche properties but the associated control parameters, and especially their accessibility in conducive functional forms e.g., pristine growth and ease of doping, especially in forms that enable incorporation and processing with standard CMOS technologies. The class of semiconducting transition metal dichalcogenides (TMDCs)\cite{splendiani2010emerging,zhu2011giant} with strain as a control parameter are particularly relevant in the above context. Strain induced modulation of band structure in atomically thin TMDCs is one of the most elegant methods for engineering their electrical, luminescence, optoelectronic and photonic properties. \cite{feng2012strain,conley2013bandgap,he2013experimental,lloyd2016band}  
Monolayer TMDCs posses incredible mechanical strength (in-plane stiffness $\sim 0.2\ TPa $), a low bending modulus ($ \epsilon \sim 9.61\ eV $) along with extreme strain endurance capability ($ \sim 10\% $),  making them ideal candidates for exploring strain engineering.\cite{bertolazzi2011stretching,peng2013outstanding,jiang2013elastic,liu2014elastic} Introduction of strain in TMDCs break the inversion symmetry, resulting in significant changes in the electronic band structure, lowering the band-gap and transitioning from direct to indirect in monolayer samples, as evidenced in the photoluminescence (PL)  properties. \cite{feng2012strain,conley2013bandgap,he2013experimental,yu2015phase}
Previous investigations have reported various straining mechanisms that includes (i) uniaxial \cite{conley2013bandgap,he2013experimental, wang2015strain}, or biaxial straining, \cite{so2021electrically,lee2022drift, lloyd2016band,frisenda2017biaxial} (ii)  draping over patterned or roughened substrates\cite{ng2022improving,huang2022large}, and through (iii) formation of wrinkles and bubbles or buckle delamination that are nucleated due to differential straining or surface adhesion\cite{deng2019strain,darlington2020imaging}.  Under tensile strain, monolayer $\textrm{MoS}_2$ shows a reduction in the optical bandgap ($E_g$) $\sim 45\textrm{--}70\ meV/\%$  under uniaxial strain \cite{conley2013bandgap,he2013experimental}, and $ \sim 100\ meV/\% $ for biaxial strain, \cite{lloyd2016band,plechinger2015control} evidencing upto $ \sim 25\% $ reduction of $E_g$.\cite{lloyd2016band} Typically, the $E_g$ changes from direct to indirect for strain beyond $ \sim 0.66\% $, \cite{conley2013bandgap,yu2015phase} which therefore allow continuous tuning of emission energy and intensities\cite{feng2012strain,lloyd2016band}. While stretching and bending of flakes mounted on flexible substrates allow for continuous, uniform and quantifiable control of \%strain and enables robust calibration of strain dependent properties, the device architecture are not conducive for simultaneous variation of other parameters like temperature, gating etc. and is not compatible with standard semiconductor device process flow.     

Conformal draping of atomically thin flakes onto  nanostructured or even roughened substrates offer an alternate straining method that alleviates many of the above shortcomings along with additional advantages, though the non-uniform nature of strain distribution makes its quantification and control more complex. Draping flakes over periodically patterned substrates, as attempted here, nucleates spatially modulated local strain and allows for a degree of control on the overall strain in the flake, via change in geometry and periodicity of the underlying pattern. In such cases, maximum strain is typically induced at the edges of the patterned nanostructures (NS), along with the wrinkles, bubbles and buckling that take place in the intervening regions between the NS, as discussed later. Such spatial localization of strain induces spatially localized modification to both the electronic and phonon band structures, where local reduction in $E_g$ creates local potential wells  in the otherwise uniform potential landscape i.e. in the conduction and valence band minima and maxima (CBM and VBM). Local reduction in $E_g$ increases the local electron density ($n_e$),\cite{chakraborty2012symmetry,chaste2018intrinsic} and under photo-excitation enhances the local PL intensity \cite{tyurnina2019strained,lee2020switchable,li2015optoelectronic}. The enhancement is aided by the ``funnelling'' effect \cite{tyurnina2019strained,castellanos2013local,lee2020switchable,darlington2020imaging}, where excitons  from surrounding regions diffuse and localize at the potential wells, which is also reported to give rise to extremely sharp  emission lines ($< 200\ \mu eV$) and create single-photon emission centres.\cite{branny2017deterministic,parto2021defect,kern2016nanoscale}
Such non-uniform strain has also been shown to increase carrier mobility ($\sim 900\ cm^{2}/Vs $) and improve device performance in mechanically exfoliated $\textrm{MoS}_2 $ monolayers.\cite{liu2019crested,ng2022improving}  The higher mobility also benefits from the associated reduction of the electron effective mass ($m_{e}^*$),\cite{dong2014theoretical} and enhanced screening due to the higher $n_e$ thus reducing scattering from charged defects. 
Further, strain induced local distortion of the atomic lattice potentially modifies the intrinsic dielectric constant ($\epsilon_D $) and the phonon density of states, adversely affecting electron-phonon scattering.\cite{ng2022improving} Together the above parameters would contribute to improvement in carrier mobility and overall electrical transport properties, as reported before.\cite{yu2015phase,yun2012thickness,harada2014computational} 	  

Here, strain induced change in local optical, electrical and electronic properties of CVD-grown monolayer $ \textrm{MoS}_2 $ flakes are investigated by transferring the flakes on to SiO$_2$/Si substrates patterned with periodic Au NS. While spatially resolved PL and Raman spectroscopy evidences the effects of local strain, bearing registry with the underlying NS pattern, the spectral red-shifts quantify the local \% strain. 
Strain induced nanoscale features such as wrinkles, nanobubbles etc. are visualized via atomic force microscopy (AFM) based measurements, which not only resolves these nanoscale features but also provides an alternate route for quantification of local strain, via local curvature of the surface.
The spatially resolved current and conductance maps from conducting AFM (CAFM) investigations directly evidences  higher $n_e$ at the strained regions by $\sim \times10$. The results are  commensurate with the observed red-shift of the $A_{1g}$ Raman mode that is sensitive to $n_e$, and corroborated by theoretical calculations that reveal the band diagram at the unstrain -- strain interface along with the accumulation of carriers in the strained regions, especially at the interfaces.  
Finally, field effect transistor (FET) measurements on these strained $\textrm{MoS}_2$ demonstrates stronger gate modulation of channel conductance, improved channel conductivity and $\times 60$ higher carrier mobility ($\mu_{FE}$), in comparison to unstrained samples. The increase in $\mu_{FE}$ is shown to scale with ``overall strain'' effected via increased NS density. 

Overall our results present a comprehensive picture of strain engineering via macroscopic patterning of substrate and the effects of induced modulation in local optical and electronic properties along with electrical transport in CVD-grown monolayer TMDCs. The wide compatibility of the methodology and ease of customization contextualizes the relevance of the present investigation, which is applicable to various fields like photonics, optoelectronics, quantum technologies etc.

\section{Results and Discussion}
\label{results}
\begin{figure}[h!]
	\includegraphics[width=15cm]{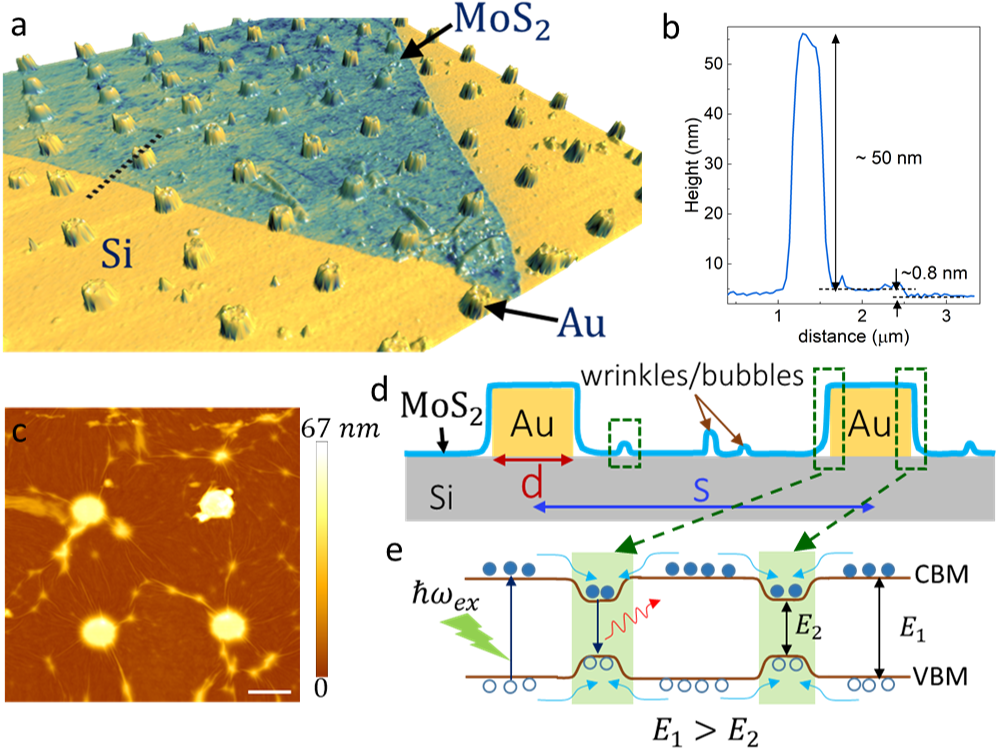}
	\centering
	\caption{(a) AFM topography of $\textrm{MoS}_2$ flake draping Au-nanostructures (NS) on Si; (b) height profile along black-dotted line in figure a; (c) topography showing wrinkles and bubbles around NS, scale bar: $500\ nm$;  (d) cross sectional schematic of  $\textrm{MoS}_2$ flake conformally draped over the NS (d: diameter $ \sim 300\ nm $, s: periodicity, h: height $ \sim 50\ nm $). Green-dashed boxes denote strained regions; (e) schematic band-diagram depicting conduction (CBM) and valence (VBM) band edges and bandgaps ($ E_1 $, $ E_2 $ ) over unstrained and strained regions, along with envisaged photoexcited charge accumulation at the strained region, denoted by curved arrows.}
	\label{fig:Figure 1}
\end{figure}
CVD-grown monolayer $\textrm{MoS}_2$ flakes were transferred onto Si substrates with pre-patterned square periodic array of Au nanocylinders of  diameter:  $d \sim 300\ nm $,  height: $ h\sim 50\ nm $ and periodicity ($s$) varying from $ 2\ \mu m $ to $ 0.5\ \mu m $. Details of patterning and sample transfer are available in Supporting Information (SI) section \ref{S1}. 
Figure \ref{fig:Figure 1}a shows the 3D topography of a monolayer  $\textrm{MoS}_2$ flake on a nanostructured substrate with $s \sim 2\ \mu m$ (see SI figure \ref{fig:si:Figure S1}). The line scan in figure \ref{fig:Figure 1}b, along the black-dotted line in figure \ref{fig:Figure 1}a, reveals the height of the cylindrical NS to be $ \sim 50\ nm $, and the monolayer thickness of $\textrm{MoS}_2$ $ \sim 0.8\ nm $. Topographically the flake conformally drapes over the Au NS, nucleating wrinkles that spread out ``radially''  from each NS along with occasional nanobubble formation in the intervening region, in-between the NS (figure \ref{fig:Figure 1}c). 
Figure \ref{fig:Figure 1}d shows a cross sectional schematic of the flake draped over the NS, with high  strain regions demarcated by dashed green lines. The presence of these local perturbations  indicate non-uniform, local strain on the flake that distorts the lattice locally, modifying the local electronic bandstructure.\cite{lloyd2016band,conley2013bandgap,feng2012strain} Figure \ref{fig:Figure 1}e shows the schematic band-diagram depicting the variation of conduction and valence band edges (CBM, VBM) across unstrained and strained regions, creating local potential wells and ``funnelling'' of photogenerated charge carriers from the surrounding regions. 

\begin{figure}[h!]	
	\includegraphics[width=16cm]{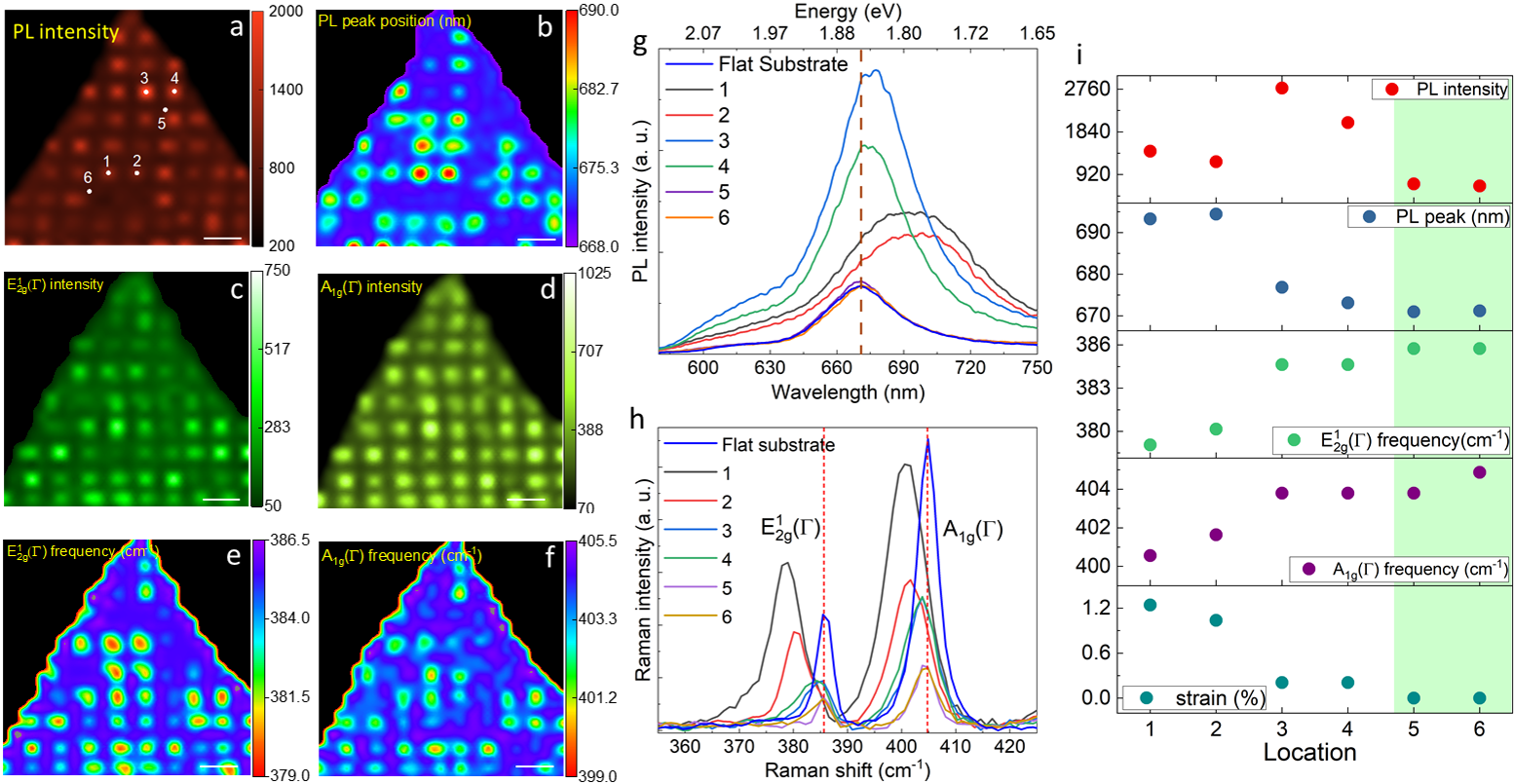}
	\centering
	\caption{Spatial maps of (a) PL intensity; (b) PL peak wavelength; (c-d) intensity and (e-f) frequency for $ E_{2g}^1(\Gamma) $ and $ A_{1g}(\Gamma) $ Raman modes of the $\textrm{MoS}_2$ flake on patterned substrate (figure \ref{fig:Figure 1}). (g) PL spectra and (h) Raman spectra at selected locations, ``on'' and ``off'' nanostructures, as marked in figure a. The vertical dashed lines indicate the spectral peak for unstrained $\textrm{MoS}_2$. (i) Correlated plots showing the variation of PL peak intensity and position, Raman mode frequencies and estimated strain from the red-shift of the $ E_{2g}^1(\Gamma)$ mode at the six points labelled in figure a. The green-patch demarcates data for ``off''-nanostructure positions. All scale bar: $ 2\ \mu m $.}
	\label{fig:Figure 2}
\end{figure} 
Figure \ref{fig:Figure 2}a shows the spatial map of the primary PL peak intensity, around $668\ nm$ corresponding to $A$ excitons\cite{mak2010atomically} across the monolayer $\textrm{MoS}_2$ on the nanostructured substrate. The map bears registry with the underlying pattern (figure \ref{fig:Figure 1}) with the peak intensity maximizing at regions around the Au NS, designated as ``on'' points (e.g. points 1, 2, 3 and 4 in figure \ref{fig:Figure 2}a), compared to the peak intensity from the ``off'' regions  lying between NS (e.g. points 5 and 6). Figure \ref{fig:Figure 2}b shows the spatial map  of the peak emission wavelength ($ \lambda_p $) that varies  between $ 668 \textrm{--} 695\ nm\ (\equiv 1.85 \textrm{--} 1.78\ eV) $ again bearing registry with the NS periodicity. $ \lambda_p $ at the ``on'' sites are red-shifted between $ 7\ nm \textrm{--}  30\ nm\ (\equiv 10 \textrm{--} 60\ meV)$ compared to that from the ``off''  points.
Individual PL spectrum acquired at the six points are shown in figure \ref{fig:Figure 2}g, along with that from an unstrained $\textrm{MoS}_2$ on a flat surface, for which the $ \lambda_p $ is denoted by the brown-dashed line. 
The higher intensity and red-shifted spectra from the ``on'' regions reflect the local reduction in bandgap of $\textrm{MoS}_2$ at the strained regions.

The locally induced strain in the $\textrm{MoS}_2$ flake is readily evidenced in spatially resolved Raman spectral analysis across the sample and evolution of the two primary modes, i.e. the in-plane $ E_{2g}^1(\Gamma) $ and out-of-plane $ A_{1g}(\Gamma) $ modes may be used to estimate the locally induced strain effectively. 
The spatial maps of the modal intensity  and frequency of the two modes are shown in figures \ref{fig:Figure 2}c and \ref{fig:Figure 2}e for  $ E_{2g}^1(\Gamma) $ and figures \ref{fig:Figure 2}d and \ref{fig:Figure 2}f for the $ A_{1g}(\Gamma) $ mode.
For either mode the Raman signal is more intense at the NS (``on'' points) that is accompanied by mode softening, with the peaks shifting to lower frequencies. 
Figure \ref{fig:Figure 2}h plots the Raman spectra acquired with $ 532\ nm $ excitation at the six  locations shown in figure \ref{fig:Figure 2}a. 
The spectra resolve the two characteristic Raman modes  at $ \sim 385.5\ cm^{-1} $ and $ \sim 405\ cm^{-1}$,\cite{lloyd2016band,li2015optoelectronic} recorded for $\textrm{MoS}_2$ on a flat surface matching with those at the ``off'' points (5 and 6) shown by red-dotted lines in figure \ref{fig:Figure 2}h. 
At the ``on'' sites (points 1 -- 4) the $ E_{2g}^1(\Gamma)$ mode shows variable  red-shift upto $ \sim 6.5\ cm^{-1} $ , which for the $ A_{1g}(\Gamma)$ mode is red-shifted upto $ \sim 3.5\ cm^{-1} $.
While the red-shift in $E_{2g}^1(\Gamma)$ arises from local strain,\cite{conley2013bandgap} 
that of $ A_{1g}(\Gamma)$ is influenced by local $n_e$ \cite{chakraborty2012symmetry} and provides a relative measure of the same. Thus, the corresponding spatial maps in figures \ref{fig:Figure 2}e and \ref{fig:Figure 2}f demarcates regions of high and low strain and $n_e$, respectively. 
The correlation between the PL and Raman maps is indicative of the common origin, i.e. strain, of either phenomena as elucidated in figure \ref{fig:Figure 2}i which co-plots the peak PL intensity, $ \lambda_p $ along with the frequency of the $ E_{2g}^1(\Gamma) $ and $ A_{1g}(\Gamma) $ modes across the six ``on-off'' locations in figure \ref{fig:Figure 2}a. 
The PL spectra at points 1 and 2 show the greatest $\lambda_p$ red-shift, commensurate with the highest red-shift of the Raman modes, signifying high strain and $n_e$. The figure also plots the strain values in the range $ 0.2\% $ to $ 1.3 \% $, estimated from the red-shift in $ E_{2g}^1(\Gamma) $ peak, following ref\cite{lloyd2016band}. 

Similarly, the spectral red-shift of the $ A_{1g}(\Gamma) $ peak (figure \ref{fig:Figure 2}f) over NS varies in the range of $ 1\ cm^{-1} $ to $ 3.5\ cm^{-1} $, indicating increased yet non-uniform local doping across the strained regions. Strain induced reduction in bandgap (red-shift $ \lambda_p $ in the PL spectrum) has been reported to be $ \sim -45\ meV/\% $ strain for uniaxial tensile strain\cite{conley2013bandgap} and $ \sim -99\ meV/\% $  under biaxial strain.\cite{lloyd2016band,plechinger2015control} Consequently, a red-shift in $ \lambda_p \sim 30\ nm (\equiv 70\ meV) $, at points 1 and 2 correspond to strain between $ 0.7 \textrm{--} 1.5\% $ depending on the uniaxial or biaxial nature of the local strain.
Similarly, PL spectra from positions 3 and 4  show less red-shift in $\lambda_p $ (bandgap lowering $ \sim15\ meV $), with lower estimated strain ($ 0.2 \% $) and high PL peak intensity enhancement ($ 3 \times$)  compared to the unstrained regions, due to the ``funnelling'' effect.
The lower PL enhancement ($1.5\times$) at the more strained points 1 and 2 is due to the more indirect nature of the bandgap, along with increased spectral broadening (figure \ref{fig:Figure 2}g), indicating  phonon-assisted indirect processes contributing to the spectra \cite{lloyd2016band,conley2013bandgap,feng2012strain,steinhoff2015efficient,niehues2018strain} (see SI section \ref{S3} for further discussion).  

Overall, these observations indicate strain-induced modification in local bandstructure of $\textrm{MoS}_2$, especially bandgap reduction at the NS, which is schematically represented in figure \ref{fig:Figure 1}e. This leads to local lowering of the CBM at the strained regions creating potential wells, spatially modulating the potential landscape, which leads to ``funnelling'' of excitons towards local potential minima (figure \ref{fig:Figure 1}e). \cite{feng2012strain,branny2017deterministic,li2015optoelectronic} As a result, strained regions in $\textrm{MoS}_2$ exhibit enhanced PL intensity in comparison to unstrained regions. CVD-grown $\textrm{MoS}_2$ monolayers are known to be $ n $-type doped and the additional lowering of the band gap at strained regions increases the probability of thermal excitation and defect ionization, thus increasing the local carrier density that is reflected in the correlated $ A_{1g}(\Gamma) $ frequency shift at the  strained regions evidenced in figures \ref{fig:Figure 2}i  and \ref{fig:Figure 2}f.

Though the spectroscopic signatures  of localized strain around the NS in the corresponding spatial maps, they lack adequate resolution to evidence strain effects on the smaller features i.e., the wrinkles and bubbles, which are well resolved in AFM results. 
\begin{figure}[h!]	
	\includegraphics[width=16cm]{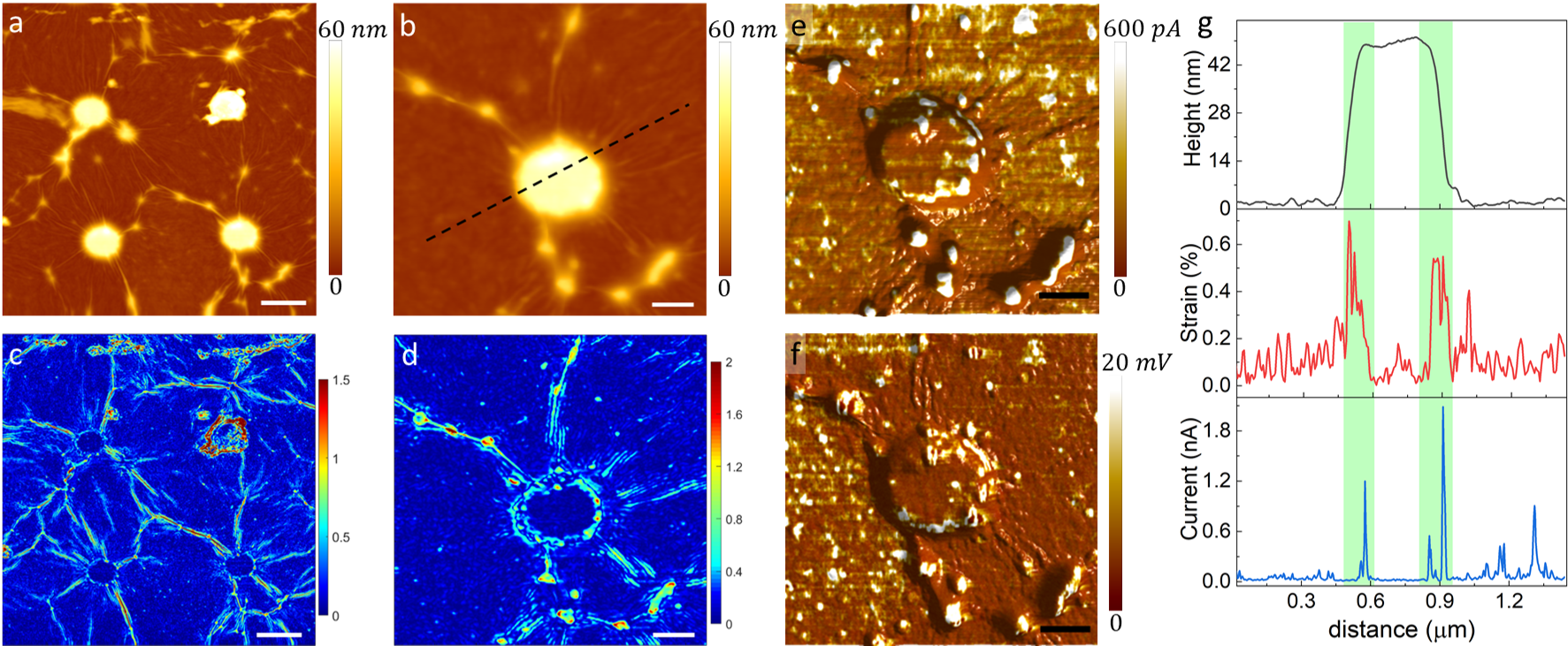}
	\centering
	\caption{(a,b) AFM topography of monolayer $\textrm{MoS}_2$ flake draped over Au-NS on Si of varying $xy$ range. (c,d) Map of strain component $ \epsilon _{zz} $ depicting strain localization calculated from topography (figure a, b). Color bar shows the variation of $ \epsilon _{zz} $ magnitude in \%. CAFM (e) current-map and (f) $ \dv{I}{V} $-map recorded simultaneously with topography, for $ +1\ V $ sample bias. (g) Correlated linescan showing the topographic height variation, strain profile and local current along the black-dashed line shown in figure b. The green shaded region demarcates the edge of a NS.  Scale bars: (a,c): $ 400\ nm $ and (b,d,e,f): $ 200\ nm $.}
	\label{fig:Figure 3}
\end{figure}         
Figures \ref{fig:Figure 3}a and \ref{fig:Figure 3}b show large and small area AFM topography of  the NS draped with an $\textrm{MoS}_2$ flake resolving the wrinkles and nanobubbles. The wrinkles,  extending ``radially'' outward from each NS are high strain features that allow relaxation of  strain in the ``flat'' regions between the NS. The features visualized with clarity in the  AFM phase images shown in figure \ref{fig:Figure S2} in SI. Figures \ref{fig:Figure 3}c, d display the map of the $ \epsilon _{zz} $ component of the strain tensor, calculated from the topographic map (figures \ref{fig:Figure 3}a, b),  using continuum elasticity theory \cite{landau1986theory} using the equation:
\begin{equation}
	\abs{\epsilon _{zz}}=\abs {\frac{\eta t}{1-\eta t} \left [\pdv[2]{h}{x}+\pdv[2]{h}{y} \right]}
\end{equation}
where, $ \eta (=0.25) $ denotes the Poisson's ratio \cite{liu2014elastic} and $ t (=0.8\ nm) $ is the thickness of monolayer $\textrm{MoS}_2$ and $ h $ represents the local topographic height variation in the sample. The $ \epsilon _{zz} $ maps (figures \ref{fig:Figure 3}c, d) not only demarcate and quantify strain in $\textrm{MoS}_2$ at the NS edges, but also accentuate the strain along the wrinkles and nanobubbles. Other intervening regions, including the top of the NS show negligible strain. The estimated \%strain lie in the range: $ 0.2\% \textrm{--}2 \% $, which is commensurate with the strain values obtained from the Raman and PL spectral shifts. Table \ref{table:si:strain} in SI tabulates and compares the estimated \%strain values from the red-shift in the PL spectra, Raman spectral shift and calculated from the AFM topography.

The effects of change in bandstructure is also reflected in the local electrical and electronic properties of the strained regions. Figures \ref{fig:Figure 3}e and \ref{fig:Figure 3}f depicts the spatially resolved conducting AFM current map and the junction $ \dv{I}{V} $ maps at $ +1\ V $ sample bias, recorded simultaneously with the topography shown in figure \ref{fig:Figure 3}b. Both the current and $ \dv{I}{V} $ maps display high tip-sample current and local conductance over the high strained regions located at the periphery of NS, over the wrinkles and at the nanobubbles, that are correlated with features in the strain map (figure \ref{fig:Figure 3}d). The correlations are better visualized in the line scans in figure \ref{fig:Figure 3}g, taken along the black-dashed line shown in figure \ref{fig:Figure 3}b. The green shaded region in figure \ref{fig:Figure 3}g denotes the edge of the NS wherein the $\textrm{MoS}_2$ is maximally strained and carries high current. Similar observations have been reported in tunnelling spectroscopy,\cite{shin2016indirect,shabani2022ultralocalized} measurements that evidence local increase in carrier density and higher conductance in strained $\textrm{MoS}_2$. This is commensurate with the observed softening of the $ A_{1g}(\Gamma) $ Raman modes at the strained regions, with higher local $n_e$. 
The current and $ \dv{I}{V} $ maps also show that the regions between the NS carry lower current and local conductance (figure \ref{fig:Figure 3}e, f). The variation is better probed  through point $ IV $ and $ \dv{I}{V} $ spectra shown in SI figures \ref{fig:si:Figure S3}b and \ref{fig:si:Figure S3}c, which are recorded at different locations marked in the accompanying topography (figure \ref{fig:si:Figure S3}a).
The $ IV $ and $ \dv{I}{V} $ spectra recorded over a nanobubble (location: 1, blue) shows higher junction current and local conductance, compared to the spectra recorded over flat region (location: 3, red). The spectra recorded  at the periphery of the strained regions (location: 2) shows the lowest current and conductance, which is indicative of local depletion in carriers. Carrier depletion at the periphery of strained regions is also evident in current map (figure \ref{fig:Figure 3}e), $ \dv{I}{V} $ map (figure \ref{fig:Figure 3}f) and the correlated linescans in figure \ref{fig:si:Figure S4} in SI, taken across multiple wrinkles, which are regions of high strain. 

The contrast in the current and conductance maps, recorded across the unstrained and strained regions of $\textrm{MoS}_2$ is better comprehended through the calculated energy band diagram  in the cross-over region. The band diagrams were calculated employing finite element method to solve the Poisson equation using COMSOL Multiphysics 5.3a. 
\begin{figure}[h!]	
	\includegraphics[width=8cm]{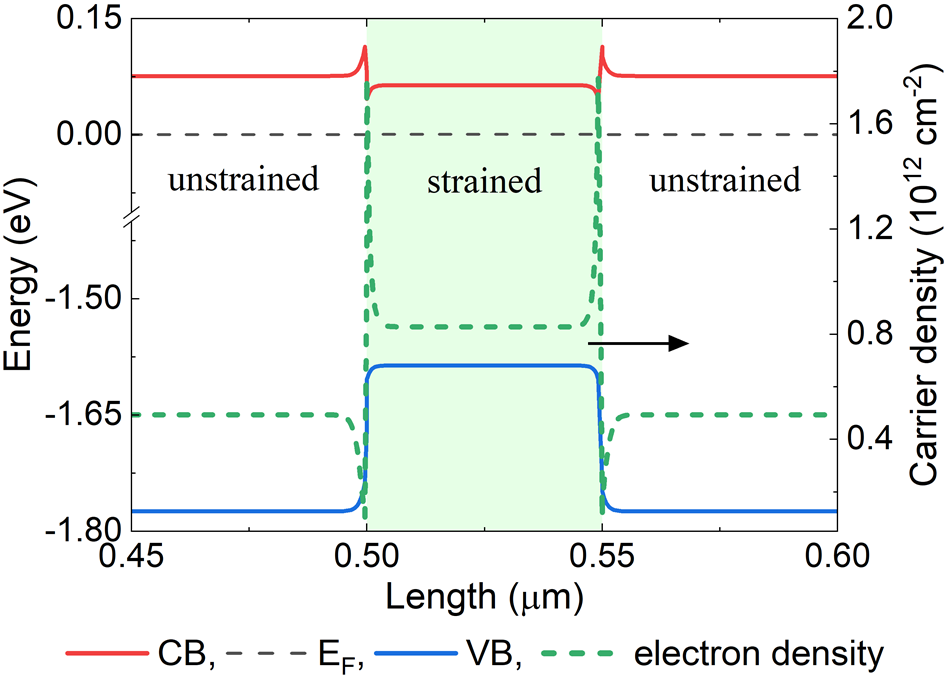}
	\centering
	\caption{Finite temperature ($ T=300\ K $), simulated band diagram and local carrier density variation along an unstrained-strained-unstrained sandwich of $\textrm{MoS}_2$ with doping density ($ n_d $) $ =10^{12}\ cm^{-2} $.}
	\label{fig:Figure 4}
\end{figure}
Figure \ref{fig:Figure 4} plots a simulated band profile for a 1-dimensional model of $\textrm{MoS}_2$, across  unstrained-strained-unstrained regions along with the variation in local electron density, calculated at $ T=300\ K $, with intrinsic $n$-type doping $ 10^{12}\ cm^{-2} $. 
SI section \ref{S8} details the geometry and physical parameters used in the model calculation. Figure \ref{fig:Figure S6} in SI plots the band diagram and variation in $n_e$ for other doping densities.  
The conduction (red) and valence (blue) band edges bend across the interfaces creating a local potential well over the strained region, with reduced band gap. 
For the conduction band edge, the interface is demarcated by a small depletion barrier of width $ <10\ nm $ and height $ \sim 38\ meV$ on the unstrained part and an accumulation region in the strained region. The calculated $n_e$ (green dashed line) varies between $ 0.8 \textrm{--} 1.8 \times 10^{12}\ cm^{-2} $ over the strained region, maximizing at accumulation region. Depletion at the peripheral unstrained regions shows a local dip in $n_e$, which  saturates $\sim 0.4 \times 10^{12}\ cm^{-2}$ away from the interface. These results are commensurate with the features of the current and conductance maps in figure \ref{fig:Figure 3} and those reported earlier. \cite{feng2012strain,li2015optoelectronic,shin2016indirect,shabani2022ultralocalized} Given the above band alignment, electron-hole pairs generated under optical illumination  will be ``funnelled'' into the strain induced potential wells thereby localizing excitons and aiding PL intensity, as reported.

Under conformal draping  of $\textrm{MoS}_2$ flakes over the NS, strain induced effects extend beyond the edge of the NS to intervening regions as wrinkles and nanobubbles, thereby making the whole flake non-uniformly strained.  Along with increase in local carrier density, non-uniform strain has been shown to affect overall electrical transport along the 2D flakes, which was investigated here by characterizing FET devices fabricated using unstrained and strained samples, in bottom gate configuration (see section \ref{S1} in SI for device architecture).
Overall strain in the samples was increased by increasing the density of the NS by reducing their spacing  as $ s: 2\ \mu m $, $ 1\ \mu m $ and $ 0.5\ \mu m $, with the corresponding FET devices  labeled as $1\times$, $2\times$ and $4\times$ strained, respectively. Higher NS density also increases the density of wrinkles and nanobubbles, as shown earlier (figures \ref{fig:Figure 1}, \ref{fig:Figure 3} and figure \ref{fig:Figure S2} in SI), which extend radially outward from each NS creating a 2-dimensional network among themselves, covering the entire flake surface, and are interspersed with occasional nanobubbles. These secondary strain features induce local strain throughout the flake. However, exact quantification of \%strain across the samples remain non-trivial due to the non-uniform nature of strain distribution. Though the local \%strain variation does not change across the samples, ``overall strain'' in the flake does increase with NS density.

\begin{figure}[h!]	
	\includegraphics[width=16 cm]{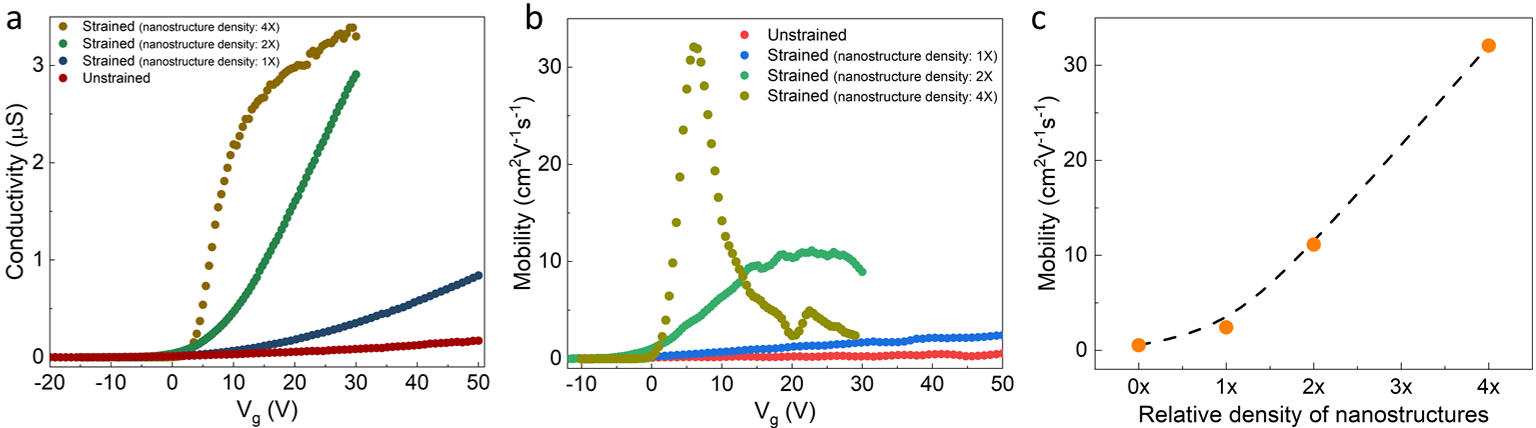}
	\centering
	\caption{(a) Two-probe transfer characteristics of strained and unstrained $\textrm{MoS}_2$. (b) Variation of field-effect mobility ($ \mu _{FE} $) with $ V_g $. (c) Variation of maximum mobility as a function of NS density.}
	\label{fig:Figure 5}
\end{figure}

\begin{figure}[h!]	
	\includegraphics[width=6cm]{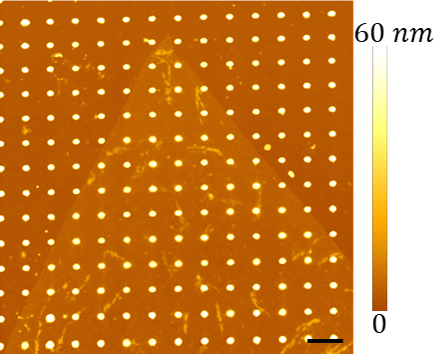}
	\centering
	\caption{AFM topography of a triangular $\textrm{MoS}_2$ flake resting on a pre-patterned Si-substrate containing Au-nanostructures of diameter: $ \sim 300\ nm $ and center to center distance: $ 1.5\ \mu m $. Scale bar: $ 2\ \mu m $.}
	\label{fig:si:Figure S1}
\end{figure}

\begin{figure}[H]	
	\includegraphics[width=14cm]{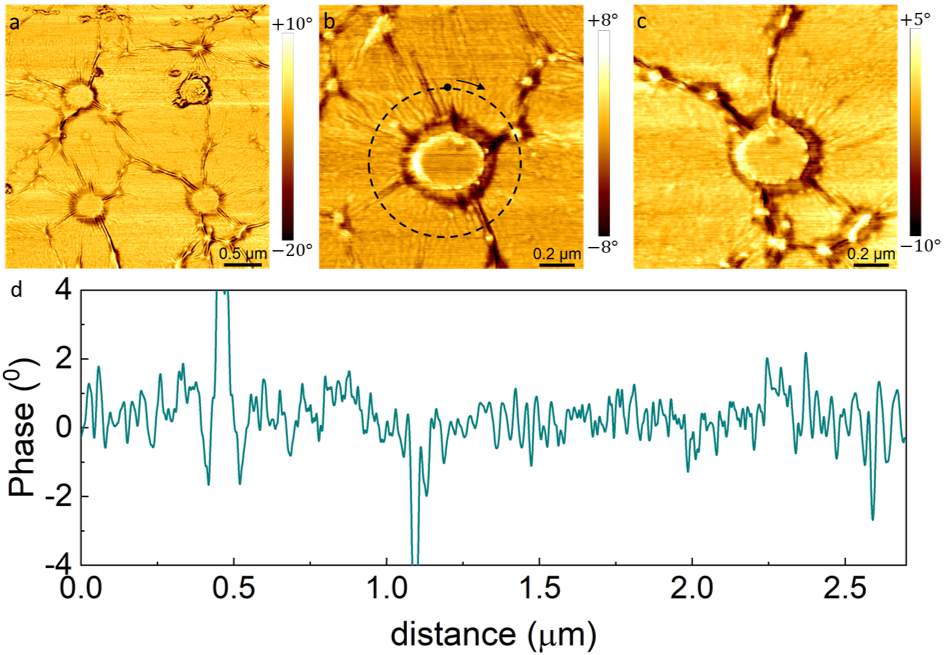}
	\centering
	\caption{(a-c)Tapping mode AFM phase maps showing wrinkles extending radially from the Au nanostructures with occasional nanobubbles across monolayer $\textrm{MoS}_2$.  (d) Line profile showing the phase-variation along the dashed-circle in figure b, where starting point of linescan represented by black-dot and the arrow indicates the direction of linescan. Existence of ripples practically all around each nanostructure is clearly evidenced.}
	\label{fig:Figure S2}
\end{figure}

\begin{figure}[H]	
	\includegraphics[width=16cm]{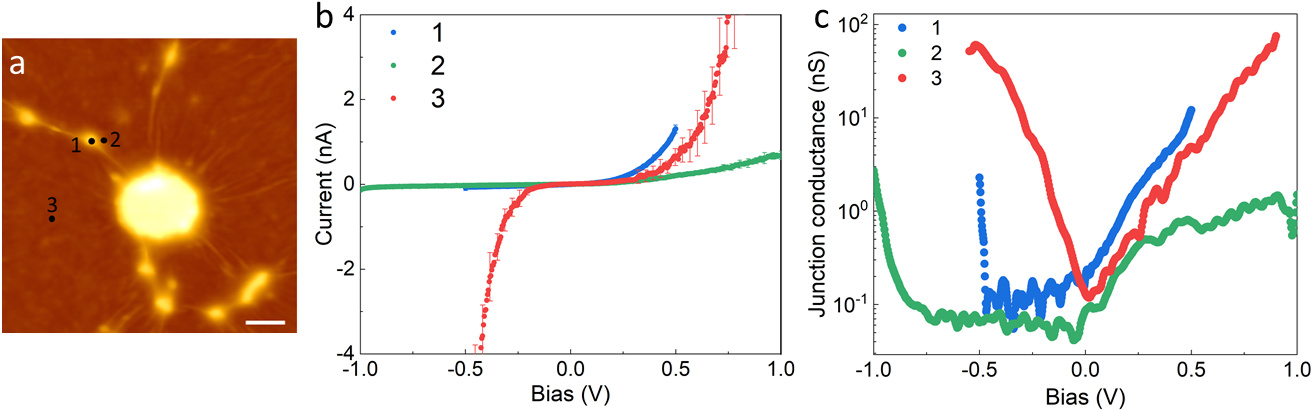}
	\centering
	\caption{(a) AFM topography of $\textrm{MoS}_2$ flake around a Au nanostructure (b) $ IV $-characteristics and (c) $ \dv{I}{V} $-characteristics at three different locations: (1) on a nanobubble , (2) in vicinity of the nanobubble and (3) far away from the nanobubble marked in figure a. Scale bar: $ 200\ nm $.}
	\label{fig:si:Figure S3}
\end{figure}

Figure \ref{fig:Figure S7} in SI shows the room temperature FET-characteristics, for the unstrained and strained ($1\times$, $2\times$, $4\times$)  devices. All the source-drain characteristics ($ I_{sd}\ vs.\ V_{sd} $) in figures \ref{fig:Figure S6}a, c, e and g exhibit ``linear'' behaviour in the $ \pm 0.2\ V $ bias range, indicating negligible impact of Schottky barrier at the contacts, at room temperature. The transfer characteristics ($ I_{sd}\ vs.\ V_g $) in figures \ref{fig:Figure S6}b, d, f and h show gate bias ($V_g$) dependent channel switching behavior, with `on-off' ratio $ >10^4 $ and exhibit $ n $-type channel conductance originating from the native point defects.\cite{zhou2013intrinsic} 
Figures \ref{fig:Figure 5}a and \ref{fig:Figure 5}b plot the variation in the two-probe channel conductivity ($\sigma_{CH}$) and field-effect mobility ($ \mu _{FE} $) calculated from the transfer characteristics, across the four devices. The plots show increasingly stronger modulation of $\sigma_{CH}$ with positive $V_g$  and enhanced $ \mu _{FE} $ with increasing strain in the $\textrm{MoS}_2$ flake, following the increase in density of nanostructures. 
Variation of $ \mu _{FE} $, calculated from the differential transconductance $ \left(\dv{I_{sd}}{V_{g}}\right) $, for each device is shown in figure \ref{fig:Figure 5}c (see SI section \ref{S10} for details). 
$ \mu _{FE} $, which  for the unstrained $\textrm{MoS}_2$ $ \sim 0.5\ cm^2V^{-1}s^{-1} $, increases to $ \sim 2\ cm^2V^{-1}s^{-1} $ for the $1 \times$ strained sample, $ \sim 11\ cm^2V^{-1}s^{-1} $ for the $2 \times$ strained sample and $ \sim 32\ cm^2V^{-1}s^{-1} $ for the $4\times$ strained sample. 

Table \ref{table:mobility} in SI section \ref{S11} summarizes the most significant reports of non-uniform strain induced $ \mu _{FE} $ enhancement reported in literature. The highest $ \mu _{FE} $ for the 4$\times$ device is lower than the highest value reported for both exfoliated  ($ \mu _{FE} = 900\ cm^{2}/Vs$)\cite{ng2022improving} and CVD-grown ($ \mu _{FE} =44.5\ cm^{2}/Vs$)\cite{huang2022large} $\textrm{MoS}_2$, it is in the ballpark of the latter. The highly non-uniform architecture employed by Huang {\it et al.} \cite{huang2022large} using underlying metal nanoparticles to locally induce strain in $\textrm{MoS}_2$ flakes precludes \%strain quantification and thus direct comparison with the present devices.  However, considering the  nanoparticle size employed and the rms roughness, it is fair to conclude that the device had significantly more strain than that in the present 4$\times$ device. Therefore it is not surprising that the maximum mobility reported in their device is slightly higher than the 4$\times$ device presented here.  Nevertheless, the present results (figure \ref{fig:Figure 5}b, c) show systematic $ \mu _{FE} $ enhancement with strain, which crosses $ \sim 60 $ for the most strained  $4\times$ device, compared to unstrained $\textrm{MoS}_2$ and is the highest enhancement reported for CVD-grown flakes and comparable to that for exfoliated ones, as evident from data in table \ref{table:mobility}.

Contrary to transport across exfoliated $\textrm{MoS}_2$ \cite{ng2022improving,radisavljevic2013mobility} which is dominated by phonon scattering near room temperature, transport in CVD-grown $\textrm{MoS}_2$, with its high native defect density, the presence of sample-substrate interfacial charge impurities and surface adsorbates, is dominated by screened Coulomb interaction mediated scattering. \cite{sze2021physics,huo2018high, yu2014towards, ghatak2011nature, lee2012synthesis}  
Importantly, CVD-grown $\textrm{MoS}_2$  samples, including the present case all show comparable $ \mu _{FE} $ that is typically $ 20\times $ smaller than that of mechanically exfoliated samples. 
Temperature dependence of $ \mu _{FE} $ is useful in delineating between phonon and charged impurity limited scattering mechanisms. 
Figure \ref{fig:Figure S8} in SI shows $ \mu _{FE}\ vs.\ T$ plots for the unstrained and $1\times$ strained devices. Both show monotonic decrease of $ \mu _{FE} $ with lowering temperature following a power law $ \mu _{FE} \propto T^p $, yielding $ p\sim +4 $. The positive value of $ p $ indicates that electrical transport in both strained and unstrained devices are primarily limited by charge-impurity scattering, within the explored $ T $ range. \cite{yu2014towards,sze2021physics,ng2022improving} 
The decrease in the free electron density, with lowering temperature weakens screening of the charged impurity potential thereby increasing electron-impurity and electron-defect scattering, which in turn suppresses $ \mu _{FE} $. Further, it has been reported that biaxial strain-induced lattice distortion leads to quenching of the phonon density of states along with an increase in the local dielectric constant, \cite{ng2022improving} and reduce the effective mass of electrons,\cite{yun2012thickness,dong2014theoretical} all of which would contribute to increase in $ \mu _{FE} $ as observed here.  

\section{Conclusions}
To summarize, we have demonstrated locally non-uniform strain-induced enhancement and control of carrier mobility in CVD-grown monolayer $\textrm{MoS}_2$ FET devices. Visualization of spatial modulation of local spectroscopic and electronic properties allow in depth comprehension of the phenomena, which shows that electrical transport in these flakes are limited by charged impurity scattering that benefits from increase in carrier density and related  effects in the differentially strained devices. Importantly, the substrate NS induced strain control scheme demonstrated here is scalable and provides a straight forward strain control parameter that is crucial for engineering device functionality, and compatible with existing on-chip processing technology for incorporation in standard electronic devices.
\section*{Acknowledgement}
Authors acknowledge Prof. M M Shaijumon and Dr. Vinayak B. Kamble (IISER Thiruvananthapuram) for use of experimental facilities. JM acknowledges financial support from SERB, Govt. of India (CRG/2019/004965), UGC-UKIERI 184-16/ 2017 (IC) and Newton Bhabha Fund, UK (IAPPI 77). AK acknowledges the PhD fellowship from IISER Thiruvananthapuram. HG and SD acknowledges DST INSPIRE for the PhD fellowship. RN is grateful to University grants commission (UGC), Government of India, for the financial support.

\section*{Authors’ contribution}
AK, SD, HG performed the experiments, SC conducted the simulations and RN did CVD-growth of samples. AK and JM wrote the manuscript with inputs from all the authors.

\section*{Notes} 
The authors declare no competing financial interest.

	\pagebreak
	\beginsupplement
	\section*{\begin{center}
		Supporting Information (SI)
\end{center}}
\section{Experimental Methods}
\label{S1}
\subsection{Substrate patterning:}
The Au-nanostructures of diameter: $ 300\ nm $,  height: $ 50\ nm $  and varying center to center distance or periodicity: $ 0.5\textrm{--}2\ \mu m $, were fabricated on $ \textrm{SiO}_2/\textrm{Si} $ and Si wafer using standard e-beam lithography pattering with PMMA as the resist using a Raith Pioneer 2 electron beam system. Metal deposition of Cr/Au ($ 5\ nm $, $ 45\ nm $) was done using a thermal evaporator, followed by standard lift-off using acetone.

\subsection{Sample growth and transfer:}
The $\textrm{MoS}_2$ samples were grown using chemical vapour deposition (CVD) technique using $\textrm{MoO}_3$ and S-powder as precursor. $ \textrm{MoO}_3 $ powder was spin-coated on $\textrm{SiO}_2$/Si substrate after dissolving in ethanol and kept inside a quartz tube and placed at the center heating zone of the furnace. $ 500\ mg $ S-powder was kept at the upstream end of the tube. Samples growth happened at $ \sim 850^\circ C $ under continuous Argon flow at $ 100\ sccm $. The as-grown $\textrm{MoS}_2$ flakes on $\textrm{SiO}_2$/Si  substrate were transferred onto various substrate by wet etching technique. First, the as-grown flakes were coated with 120K PMMA (dissolved in anisole at $ 10\ wt\% $), followed by baking at $ 140^\circ C $ for $ 15 $ mins. The coated substrates were floated on 2M aqueous NaOH solution at $ 60^\circ C $, to etch out the $\textrm{SiO}_2$ layer and detach the flakes from substrate into the PMMA film. Later, the detached film was scooped and transferred into water bath for $ 3 $-cycles with a duration of $ 30 $ mins for each. The film was transferred by scooping it from water bath with the relevant substrate and subsequently baked at $ 80^\circ C $ for $ 2 $ hr to evaporate the water. Finally, the flakes were obtained after dissolving the PMMA layer in warm acetone, followed by IPA wash and blow dried with air.

\subsection{Spectroscopic study:}
The spatially resolved photo-luminescence and Raman studies were conducted using a HORIBA Xplora Plus Raman setup based on a confocal microscope. Raman mapping and spectra acquisition were done with $ 532\ nm $ laser excited through a $ 100X $ objective with NA: $ 0.9 $. Spectra for PL and Raman were recorded with TE-cooled ($ -60^\circ C $) CCD, using grating with ruling density of $ 600\ gr/mm $, and $ 2400\ gr/mm $, respectively. All the spectroscopic measurements were conducted in room temperature.  

\subsection{Atomic Force Microscopy investigations:}
All the AFM investigations were conducted using a Bruker Multimode 8 AFM under ambient conditions. The conducting AFM current maps and $ \dv{I}{V} $-maps were recorded using the TUNA 2 pre-amplifier, having variable voltage gain and capable of detecting current in the range of $ 1\ fA \textrm{--} 1\ \mu A  $. Both the current-map and $ \dv{I}{V} $-maps were recorded simultaneously with topography using gold-coated probes (HQ:CSC37/Cr-Au, $ k = 0.1 \textrm{--} 0.6\ N/m $) from MikroMasch. For all CAFM maps, bias was applied to the sample, while probe was kept at virtual ground potential.   

\subsection{Field Effect Transistor Device Fabrication and Transport measurements:}
The FET devices were fabricated on $ n^{++} $-Si substrate coated with $ 285\ nm $ thick, thermally grown $\textrm{SiO}_2$ layer, which served as the gate dielectric. The source-drain contacts were patterned using e-beam lithography, followed by thermal deposition of Cr/Au ($ 5\ nm/30\ nm $). The transistor characteristics were measured using KEYSIGHT B2902B dual channel source-meter. The low-temperature measurements were conducted inside a closed-cycle He-cryostat (ARS Inc.).

\section{AFM topography of transferred $\textrm{MoS}_2$ flake}
\label{S2} 
\section{Discussion on change in PL characteristics with strain}
\label{S3}
Previous investigations show that the momentum and energy of the conduction and valence band minima (CBM, VBM), which are located at the $ K(K') $ points in the $E-k$ space for monolayer $\textrm{MoS}_2$, changes as a function of tensile strain and the nature of the bandgap evolves from direct to indirect.\cite{conley2013bandgap,feng2012strain,steinhoff2015efficient} According to Steinhoff \textit{et al}. \cite{steinhoff2015efficient}, smaller values of biaxial tensile strain ($ <0.6 \% $) retains the direct nature of the bandgap while it's magnitude reduces due to downshift of the CBM and unchanged VBM at $ K(K') $. Experimentally, this results in an increase in PL intensity with commensurate red-shift in emission peak. However, for higher tensile strain ($ >1 \% $), the VB energy at $ \Gamma $-point increases above the its value at the $ K(K') $-point, consequently the bandgap becomes indirect and the optical transitions are dominated by phonon-assisted processes, which in turn reduces the PL intensity. The initial  rise in PL intensity under tensile strain is then understood to arise from  nonequilibrium distribution of carriers predominantly in the $ K $-valleys, which leads to strong emission from the $ A $-exciton decay. The results presented here in figure \ref{fig:Figure 2} show similar behavior, PL spectra from positions 3 and 4, with lower strain ($ 0.2 \% $) show less red-shift ($ \sim15\ meV $) and highest PL peak intensity. The higher strained regions (points 1 and 2) with strain $ \sim 1.2 \% $ show $ \lambda_p $ red-shifted by $ \sim 60\ meV $ with reduced intensity due to the more indirect nature of the bandgap \cite{lloyd2016band,conley2013bandgap,feng2012strain}, along with increased spectral broadening, indicating  phonon-assisted indirect processes contributing to the spectra. \cite{steinhoff2015efficient,niehues2018strain} 

\section{AFM phase map}
\label{S4}	
\section{Local $ IV $ and $ \dv{I}{V} $-characteristics}
\label{S5} 
Comparison of the  $ IV $ and $ \dv{I}{V} $-characteristics across the three points shows that the strained region (point 1) carries the highest current and shows highest $ \dv{I}{V} $, whereas the immediately adjoining regions at the edge of the strained region (point 2) shows the lowest values, under positive bias. Regions away from the regions of high strain (point 3) show intermediate values.

\section{Correlated Linescans}
\label{S6} 
\begin{figure}[H]	
	\includegraphics[width=16cm]{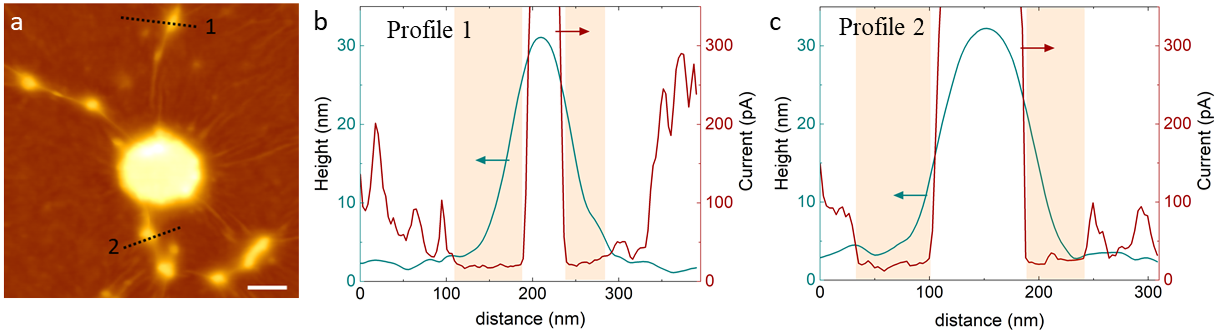}
	\centering
	\caption{(a) AFM topography of $\textrm{MoS}_2$ flake around a Au nanostructure (b, c) Correlated line profile showing the variation of CAFM current and height along the black dotted lines over the nanobubbles, as depicted in figure a. }
	\label{fig:si:Figure S4}
\end{figure}

The correlated line scans in figures \ref{fig:si:Figure S4}b and c, between topography and current maps show that the current at the top of high strain region at the nanobubble is the highest (saturated for the current amplifier gain used).The color-shaded regions in the plots demarcate the lowest current carrying regions that surround the high strain nanobubble region. This low current in these surrounding regions denotes local carrier depletion, arises due to strain induced bandgap modulation and band-bending across the interface of strained-unstrained $\textrm{MoS}_2$. 

\section{Comparison of estimated strain from PL, Raman and AFM study}
\label{S7}	
\begin{table}[H]
	\caption{Estimated strain from PL, Raman spectra and AFM topography}
	\centering
	\begin{tabular}{l c c}
		\hline\hline
		Probe & Observed shift & Estimated strain (\%)\\
		\hline\hline
		PL& $\Delta E_g (15\ \textrm{--}\ 60\ meV)$ & $ 0.2 \textrm{--} 1.5 $ \\
		Raman& $\Delta \omega (E_{2g}^1(\Gamma)) (1.0\ \textrm{--}\ 6.5\ cm^{-1})$ & $ 0.2 \textrm{--} 1.3 $ \\
		AFM & $\Delta h(x, y) $ & $ 0.4\ \textrm{--}\ 2.0 $ \\
		\hline
	\end{tabular}
	\label{table:si:strain}
\end{table}

\section{Simulating the energy band diagram across strained-unstrained interface}
\label{S8}
To numerically calculate the band profile at the interfaces of alternating unstrained-strained-unstrained regions of $\textrm{MoS}_2$, finite element method modelling was performed using COMSOL Multiphysics 5.3a. A 1D simulation domain as shown in figure \ref{fig:Figure S6} was divided into 3 regions, with the middle strained region of $ 50\ nm $ width and under 2\% strain, while the unstrained regions were taken to be $ 500\ nm $ wide.
\begin{figure}[H]	
	\includegraphics[width=6cm]{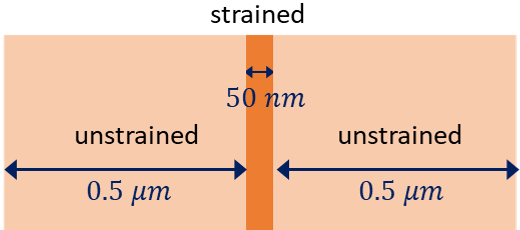}
	\centering
	\caption{Schematic of simulation domain of a unstrained-strained-unstrained $\textrm{MoS}_2$.}
	\label{fig:Figure S5}
\end{figure}
The physical parameters used to model the strained and unstrained regions are tabulated in table \ref{table:parameters}.
\begin{table}[h!]
	\caption{Physical paremeters of $ \textrm{MoS}_2 $ used for the simulation of band diagram}
	\centering
	\begin{tabular}{l c c}
		\hline\hline
		property & unstrained & strained\\
		\hline\hline
		width (nm) & $ 500 $ & $ 50 $\\
		relative permittivity & $ 4.8 $ \cite{kumar2012tunable} & $ 4.8 $\\
		Strain (\%) & $ 0 $ & $ 2 $\\
		Bandgap (eV)& $ 1.85 $ & $ 1.65 $ \\
		Electron affinity (eV) & $ 4.27 $ \cite{zhang2016systematic} & $ 4.37 $\\
		$m_e^*/m_e$ effective mass & $ 0.48 $ \cite{yun2012thickness} & $ 0.45 $\\
		doping concentration ($ cm^{-2} $)& \multicolumn{2}{c}{$ 0 $, $ 10^{10} $, $ 10^{11} $, $ 10^{12} $}\\
		dopant ionization energy (eV) & $ 0.065 $ & $ 0.056 $ \\
		\hline
	\end{tabular}
	\label{table:parameters}
\end{table}	
All calculations were performed at finite temperature $ T=300\ K $ and no variation to background relative permittivity due to strain was incorporated.
Following the $ k.p $ perturbation  theory \cite{peter2010fundamentals} the effective mass of the unstrained and strained regions are obtained as $m_e^*/m_e = 0.48 $ and $ 0.45 $, respectively \cite{yun2012thickness,dong2014theoretical} which also affects the ionization energies of the dopant states which are calculated employing the ``hydrogenic impurity model'' \cite{sze2021physics}.  The energy band diagrams for the composite $\textrm{MoS}_2$ system for various dopant densities ($n_d$) are shown in the figure below. Previous reports show that $\textrm{MoS}_2$ is $n$-type doped with typical electron density $\sim 10^{10} - 10^{12}/cm^2$.\cite{qiu2013hopping}
\begin{figure}[H]	
	\includegraphics[width=16cm]{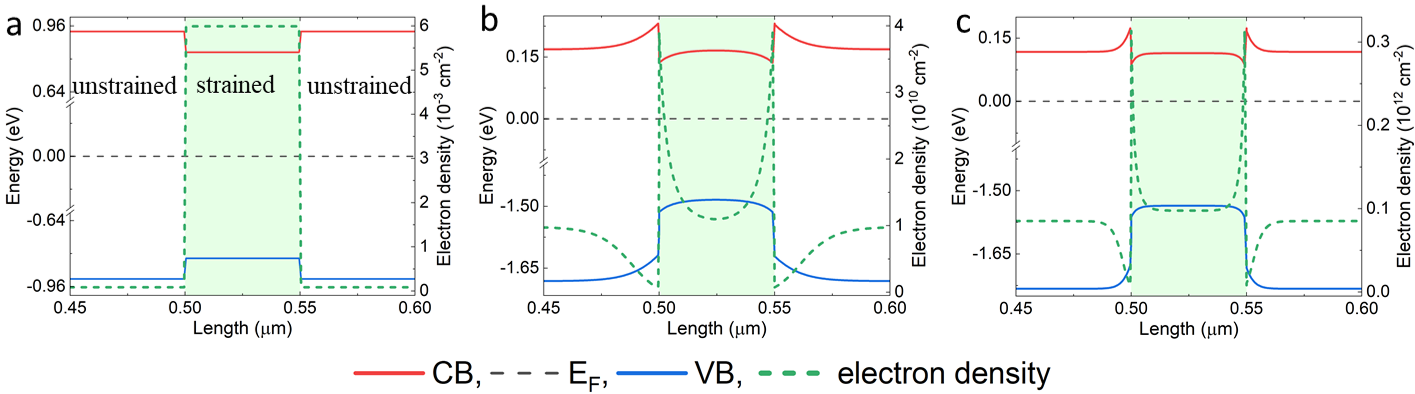}
	\centering
	\caption{Simulated band diagram and electron density  variation across an unstrained-strained-unstrained regions of $\textrm{MoS}_2$ for various doping density ($n_d$): (a) $n_d=0$, (b) $n_d=10^{10}/cm^{2}$, (c) $n_d=10^{11}/cm^{2}$.}
	\label{fig:Figure S6}
\end{figure}

\section{FET characteristics}
\label{S9}	
\begin{figure}[H]	
	\includegraphics[width=12cm]{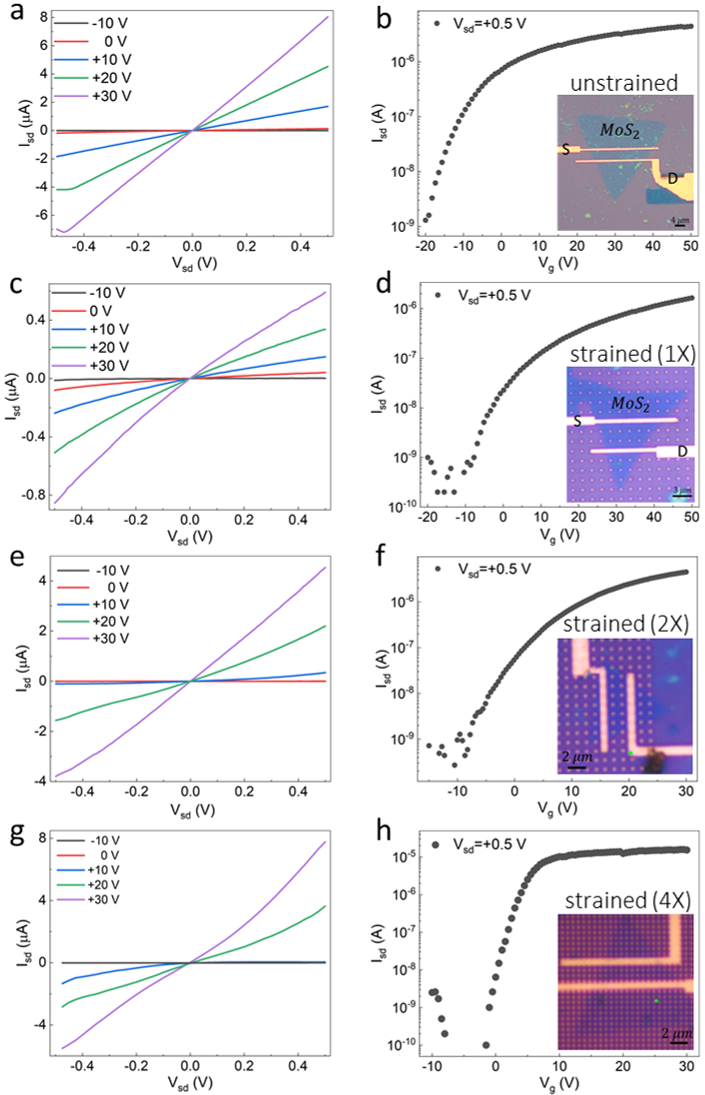}
	\centering
	\caption[FET characteristics of unstrained and strained monolayer $\textrm{MoS}_2$ devices]{Room temperature FET characteristics for (a-b) unstrained  and (c-h) variously strained monolayer $ \textrm{MoS}_2 $ flakes on $\textrm{SiO}_2$/Si. Figures on the left show source-drain characteristics as a function of gate bias $ V_g $ and on the right the figures show the transfer characteristics ($I_{sd}\ vs.\ V_g $) for $ V_{sd}=+0.5\ V $. The insets show optical images of the corresponding devices along with the Au source-drain electrodes.}
	\label{fig:Figure S7}
\end{figure}

\section{Mobility Calculation}
\label{S10}
Field-effect mobility was calculated using the formula:
\begin{equation}
	\mu _{FE}=\frac{L}{W} \times \frac{t_{ox}}{\epsilon _0 \epsilon _r V_{sd}} \times \dv{I_{sd}}{V_{g}}
\end{equation}
where, $ t_{ox} $ is oxide layer thickness, $ \epsilon _0 $ denotes vacuum permittivity, $ \epsilon _r (=3.9)$  is dielectric constant of $\textrm{SiO}_2$, and $ L $, $ W $ are channel length and width, respectively.
The $ \mu _{FE} $ values were calculated from two-probe FET characteristics and are known to be limited by the nature of the electrical contacts. High contact resistance values lead to underestimating the value of $ \mu _{FE} $. \cite{chang2014mobility} 	The linear nature of the  source-drain characteristics in the present devices suggest minimal influence of contact resistance in the estimated mobility values. \cite{nasr2019mobility}

\section{Comparison of mobility enhancement with strain}
\label{S11}
\begin{table}[H]
	\caption{Variation of mobility with strain in published literature and present investigation:}
	\centering
	\begin{tabular}{l c c c c}
		\hline\hline
		{\small sample} & {\small strain} & {\small $ \mu _{FE} (cm^{2}/Vs)$} & $ \frac{\mu _{FE} (strained)}{\mu _{FE}(unstrained)} $ & {\small reference}\\
		\hline\hline
		{\small 1L $\textrm{MoS}_2$ (CVD)}  & {\small uniform tensile $ 0.7\% $} & {\small $ 10.8 $} & {\small $ 2 $} & {\small \cite{datye2022strain}}\\
		
		{\small 1L $\textrm{MoS}_2$ (CVD)}  & {\small non-uniform $ 1.2\% $ (max)} & {\small $ 44.5 $} & {\small $ 23 $} & {\small \cite{huang2022large}}\\
		
		{\small 1L $\textrm{MoS}_2$ (CVD)} & {\small non-uniform $ 0.2 \textrm{-} 2.0\% $} & {\small $ 32 $} & {\small $ 60 $} & {\small present work}\\
		
		{\small 1L $\textrm{MoS}_2$ (exfoliated)} & {\small non-uniform $ 0.1 \textrm{-} 6.0\% $} & {\small $ 448 $} & {\small $ 50 $} & {\small \cite{ng2022improving}}\\
		
		{\small 2L $\textrm{MoS}_2$ (exfoliated)} & {\small non-uniform $ 0.1 \textrm{-} 6.0\% $} & {\small $ 900 $} & {\small $ 100 $} & {\small \cite{ng2022improving}}\\
		
		{\small 1L $\textrm{MoS}_2$ (exfoliated)} & {\small non-uniform $ 0.7 \textrm{-} 3.5\% $} & {\small $ 850 $} & {\small$ 100 $} & {\small \cite{liu2019crested}}\\
		
		{\small 1L $\textrm{MoSe}_2$ (exfoliated)} & {\small non-uniform $ 0.7 \textrm{-} 3.5\% $} & {\small $ 285 $} & {\small $ 44 $} & {\small \cite{liu2019crested}}\\
		
		{\small 1L $\textrm{WSe}_2$ (exfoliated)} & {\small non-uniform $ 0.7 \textrm{-} 3.5\% $} & {\small $ 158 $} & {\small $ 10 $} & {\small \cite{liu2019crested}}\\
		\hline
	\end{tabular}
	\label{table:mobility}
\end{table} 

\section{Temperature-dependent mobility}
\label{S12}
\begin{figure}[H]	
	\includegraphics[width=8 cm]{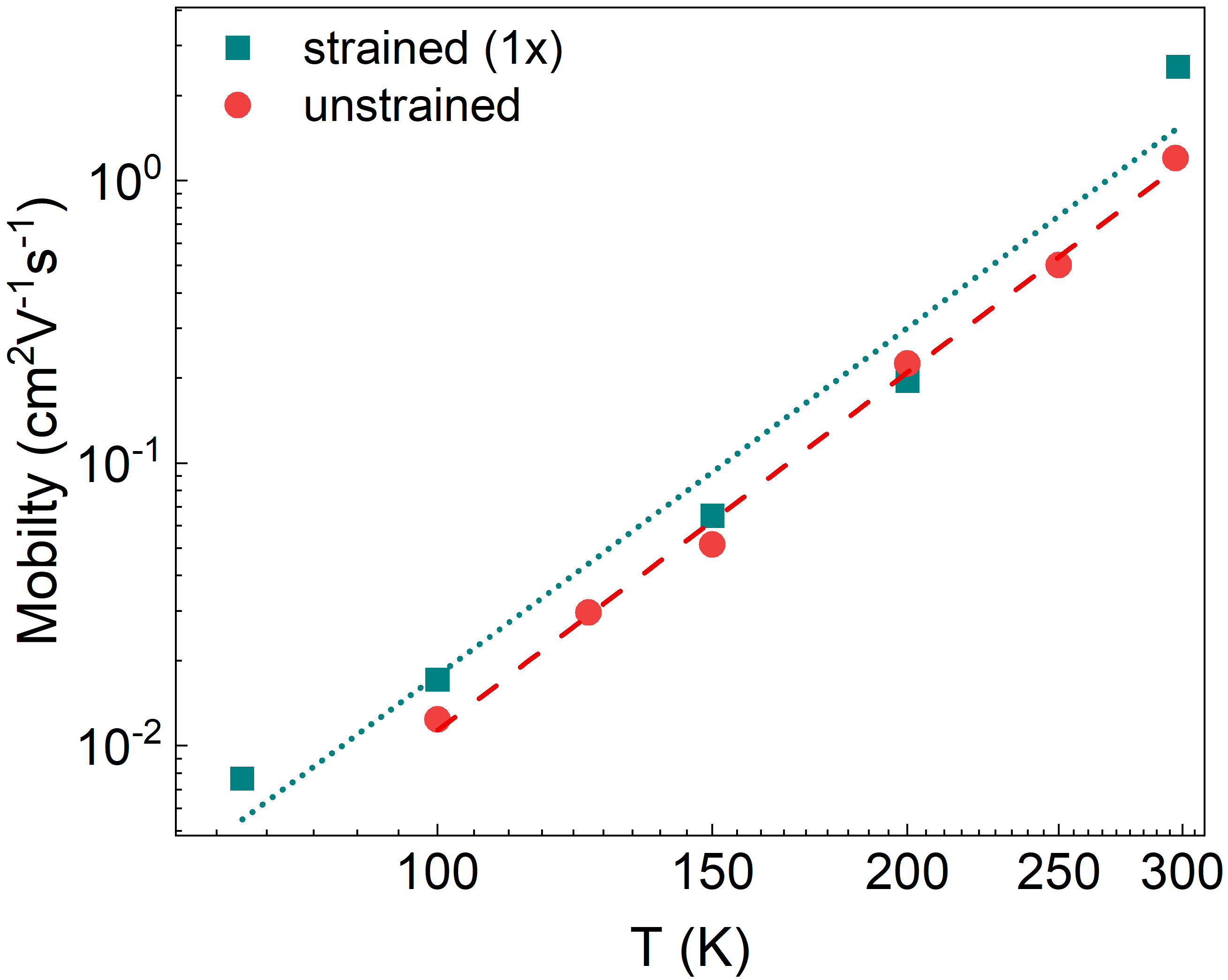}
	\centering
	\caption[Temperature dependent transport]{Mobility vs. Temperature for unstrained and strained(1x) $\textrm{MoS}_2$ with a linear fit to the equation: $\mu_{FE} \propto T^p$, where $p=4.08$ for strained (1x) and $4.21$ for unstrained sample.}
	\label{fig:Figure S8}
\end{figure}

	\printbibliography
\end{document}